# Magnetic fields alter tunneling in strong-field ionization

A. Hartung[1]*, S. Eckart[1], S. Brennecke[2], J. Rist[1], D. Trabert[1], K. Fehre[1], M. Richter[1], H. Sann[1], S. Zeller[1], K. Henrichs[1], G. Kastirke[1], J. Hoehl[1], A. Kalinin[1], M. S. Schöffler[1], T. Jahnke[1], L. Ph. H. Schmidt[1], M. Lein[2], M. Kunitski[1], R. Dörner[1]*

[1] Institut für Kernphysik, Goethe Universität Frankfurt, Max-von-Laue-Straße 1 D-60438 Frankfurt am Main, Germany

[2] Institut für Theoretische Physik, Leibniz Universität Hannover, Appelstr. 2, 30167 Hannover, Germany

* Correspondence to: (A.H.) hartung@atom.uni-frankfurt.de & (R.D.) doerner@atom.uni-frankfurt.de

**Introductory paragraph**

When a strong laser pulse induces the ionization of an atom, momentum conservation dictates that the absorbed photons transfer their momentum $p_\gamma = E_\gamma/c$ to the electron and its parent ion. Even after 30 years of studying strong-field ionization, the sharing of the photon momentum between the two particles and its underlying mechanism are still under debate in theory[1–4]. Corresponding experiments are very challenging due to the extremely small photon momentum (~$10^{-4}$ a.u.) and their precision has been too limited, so far, to ultimately resolve the debate.[5–7] Here, by utilizing a novel experimental approach of two counter-propagating laser pulses, we present a detailed study on the effects of the photon momentum in strong-field ionization. The high precision and self-referencing of the method allows to unambiguously demonstrate the action of the light's magnetic field on the electron while it is under the tunnel barrier, confirming theoretical predictions[1–3,8], disproving others[5,9,10]. Our results deepen the understanding of, for example, molecular imaging[11,12] and time-resolved photoelectron holography[13].

**Main Text**

The advent of pulsed laser systems, which are capable of generating ultrashort light pulses with electric field amplitudes on the order of the atomic binding field, launched the research field of strong-field ionization. The overwhelming majority of theoretical studies in this discipline consider ionization within the "electric dipole approximation", which neglects the linear momentum of the photon $p_\gamma = E_\gamma/c$. Without the radiation pressure of the laser light, the momentum distribution of the emerging electron-ion system is by definition symmetric with respect to the propagation direction of the light. In the wave picture, the dipole approximation disregards effects of the light's magnetic field and of the spatial inhomogeneity of the electromagnetic wave. In most cases, the dipole approximation is appropriate, as the photon momentum is typically 3-4 orders of magnitude smaller than the momenta of emitted photoelectrons and ions. However, any rigorous approach needs to account for momentum conservation as one of the most essential concepts in physics. A comprehensive understanding of the role of the photon momentum in strong-field ionization is therefore of fundamental interest.

Within a wide range of laser intensities, strong-field ionization occurs as tunneling of the bound electron through a potential barrier created by the superposition of the atomic potential and the electric field of the laser. The potential role of the light's magnetic field in such tunnel ionization processes implies interesting questions, for example, whether particles are susceptible to magnetic fields during tunneling. Apart from fundamental interest, a complete modelling – including the photon momentum – can help to better understand strong-field phenomena like high-harmonic generation[14–16] (and thus attosecond science[17,18]), time-resolved photoelectron holography[13] or molecular imaging[11,12], which are all sensitive to small perturbations of the light field.

The energy $E_\gamma$ provided by the laser field for an ionization process is accompanied by an injection of corresponding linear momentum $p_\gamma = E_\gamma/c$ to the electron-ion-system. The energy $E_\gamma$ is employed to overcome the ionization potential $I_p$ as well as to give the emitted electron its kinetic energy $E_{e,kin}$. The ionization event can be viewed as a two-step process. In a first step, the ionization potential $I_p$ must be overcome. One might thus expect that the corresponding photon momentum is transferred to the center-of-mass of the system, i.e. essentially to the much heavier parent ion (equation (1), left, with x being the direction of light). If in the second step the electron is accelerated by the laser field independently of its parent ion, one might expect that it absorbs the photon momentum associated to its gain in kinetic energy (equation (1), right).

$$< P_{ion,x} > = \frac{I_p}{c} \qquad\qquad < P_{e,x} > = \frac{E_{e,kin}}{c} = \frac{P_e^2}{2c} \qquad (1)$$

In recent work, Chelkowski et al.[1] predicted a surprising deviation from this intuitive consideration: The photon momentum associated to $I_p$ is not solely given to the parent ion, but a substantial fraction, evaluated as $1/3$ of $I_p/c$ within relativistic tunneling theory[2,8], is imparted on the electron. This offset is induced by the action of the laser magnetic field on the electron, while it performs its purely quantum mechanical motion under the tunnel barrier.[2,8] Lately, numerical solutions of the non-dipole time-dependent Schrödinger equation (TDSE) in 2D and 3D confirmed that prediction for circularly polarized light and theory found further details for the case of linear light.[4,19–23] To summarize, by putting in the recent theoretical work, equation (1) stemming from an intuitive classical perspective can be expanded to equation (2).

$$< P_{e,x} > = \frac{E_{e,kin}}{c} + \frac{1}{3} \cdot \frac{I_p}{c} = \frac{P_e^2}{2c} + \frac{1}{3} \cdot \frac{I_p}{c} \qquad (2)$$

So far, there is no experimental observation of the two non-dipole effects suggested above, namely the quantum mechanical $I_p$-dependent shift and the dependence on the overall electron momentum independent of laser intensity. This undertaking is extremely challenging, mainly because the expected offset momentum is orders of magnitude smaller than the typical momentum of the electron. In a pioneering attempt of measuring non-dipole effects, Smeenk et al.[5] found a forward shift of the electron's average momentum in circularly polarized light on the order of $10^{-2}$ a.u., increasing linearly with laser intensity. Ludwig et al.[6] and later on Maurer et al.[7] examined the peak of electron distributions in linearly and elliptically polarized light. They detected a counter-intuitive shift in the direction opposite to the laser propagation on the same magnitude as Smeenk et al. Due to the experimental resolution and the used calibration method for finding the zero of the momentum distribution, the precision of previous experimental studies did not allow to address both summands in equation (2) (see methods for more details). For the present work, we employed a novel experimental approach based on two counter-propagating laser pulses.

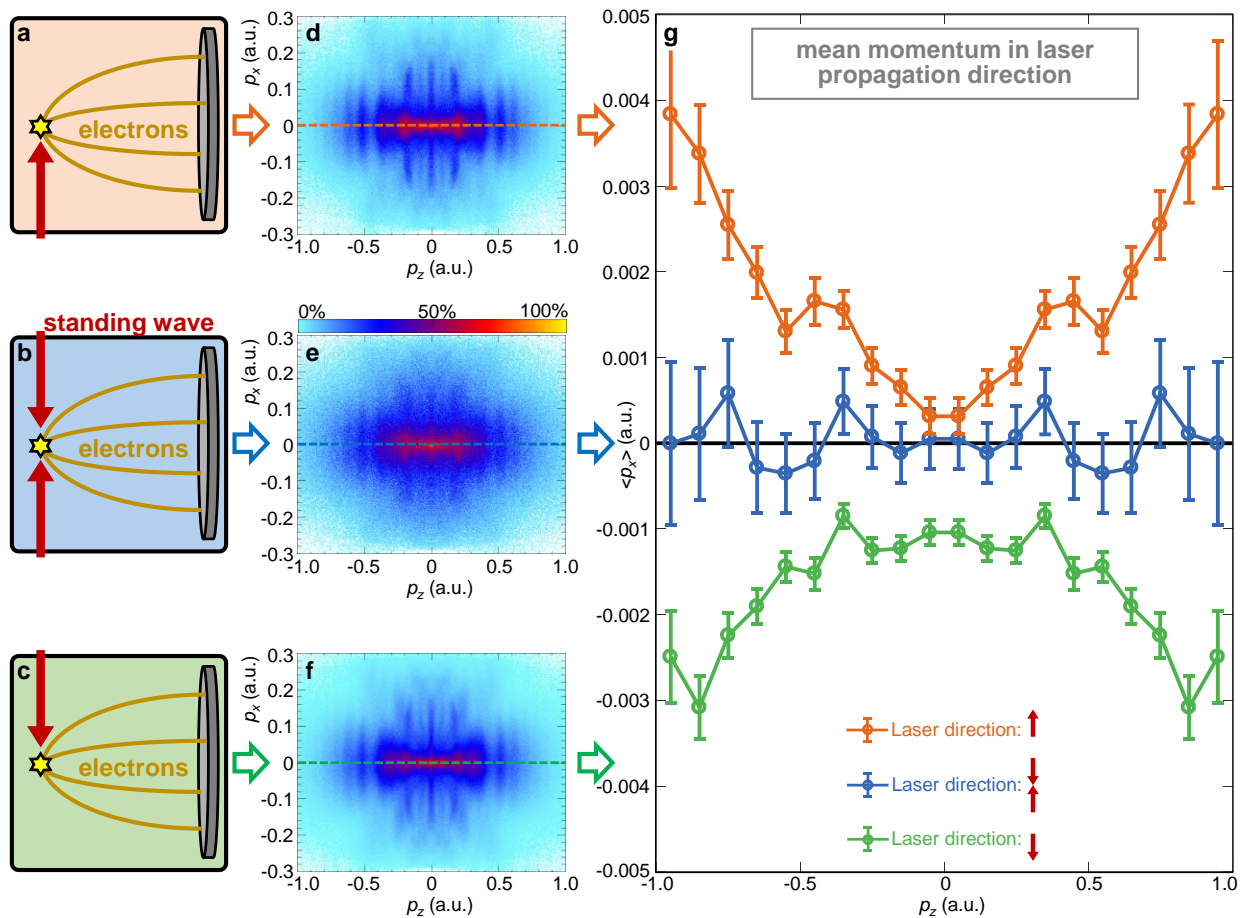

**Figure 1 | Experimental scheme.** In all panels the abscissa depicts the light polarization axis z, the ordinate depicts the light propagation axis x. **a-c**, The setup allows for three different geometries for strong-field ionization of argon. The laser can enter the interaction chamber either from one side individually (**a, c**) or from both sides simultaneously, creating a standing wave (**b**). **d-f**, The corresponding electron momentum distributions obtained for linear light do not seem to exhibit any offsets caused by the photon momentum, i.e. not showing any asymmetrical features along light propagation axis. Because of the symmetry of the experiment in light polarization axis z, the data was symmetrized in that dimension. **d**, Total counts in figure is $2.3\times10^6$, in **e** $1.0\times10^6$ counts and in **f** $5.1\times10^6$ counts. **g**, By calculating the mean momentum in light propagation direction for all three ionization scenarios and zooming in by a factor of ~100, clear non-dipole features become visible. In the case of ionization by a standing wave, no significant momentum offset arises, because of the spatial symmetry of the experimental arrangement. In the two cases of ionization by a single laser pulse from just one side, the parabolic shape flips sign when flipping the propagation direction of the laser pulse. The error bars show statistical errors.

We utilized the COLTRIMS technique[24] to measure the momentum distributions of electrons and ions created by strong-field ionization of argon in a 25 fs laser pulse with a central wavelength of 800 nm. As can be seen in Fig. 1, our setup allows for the simultaneous injection of the laser beam from opposite directions. This gives rise to three possible schemes to induce strong-field ionization: shooting in from the bottom (Fig. 1a), shooting in from the top (Fig. 1c) and creating a standing wave of light in the interaction region by simultaneous injection of light from both sides (Fig. 1b). This scheme provides essential benefits for experimental investigation of the non-dipole effects. Firstly, ionization in a standing wave (see Fig. 1b) does not – by definition – exhibit any forward-backward asymmetry, providing an intrinsically known zero momentum. Secondly, by comparing the two momentum distributions measured by injecting single laser beams from either side, instrumental asymmetries along the propagation axis of the laser cancel out.

The momentum distributions recorded separately are shown in Fig. 1d-f. The momentum of a single photon at 800 nm is $4 \times 10^{-4}$ a.u. Accordingly, obvious differences cannot be expected in the panoramic views of the electron momentum distributions in the range ±1.0 a.u (Fig. 1d-f). Hence, for better visualization the mean momentum $\langle p_x \rangle$ in direction of light is taken for every value of momentum in polarization direction $p_z$. The comparison of the three possible laser irradiation schemes in Fig. 1g clearly shows an effect caused by the photon momentum. As expected, we find $\langle p_x \rangle$ close to zero for the standing wave measurement, lying symmetrically in between the distributions belonging to the two "single-way" experiments. These exhibit mirror-symmetrical momentum offsets along their respective directions of light propagation and show an approximately quadratic dependence on the momentum, as suggested by equation (2). Fig. 1 illustrates the experimental procedure and shows data measured with linearly polarized light. Here strong post-ionization Coulomb interaction may alter the pure non-dipole concept introduced above (equation (2)). A simpler case of recollision-free ionization can be examined by using circularly polarized laser light.

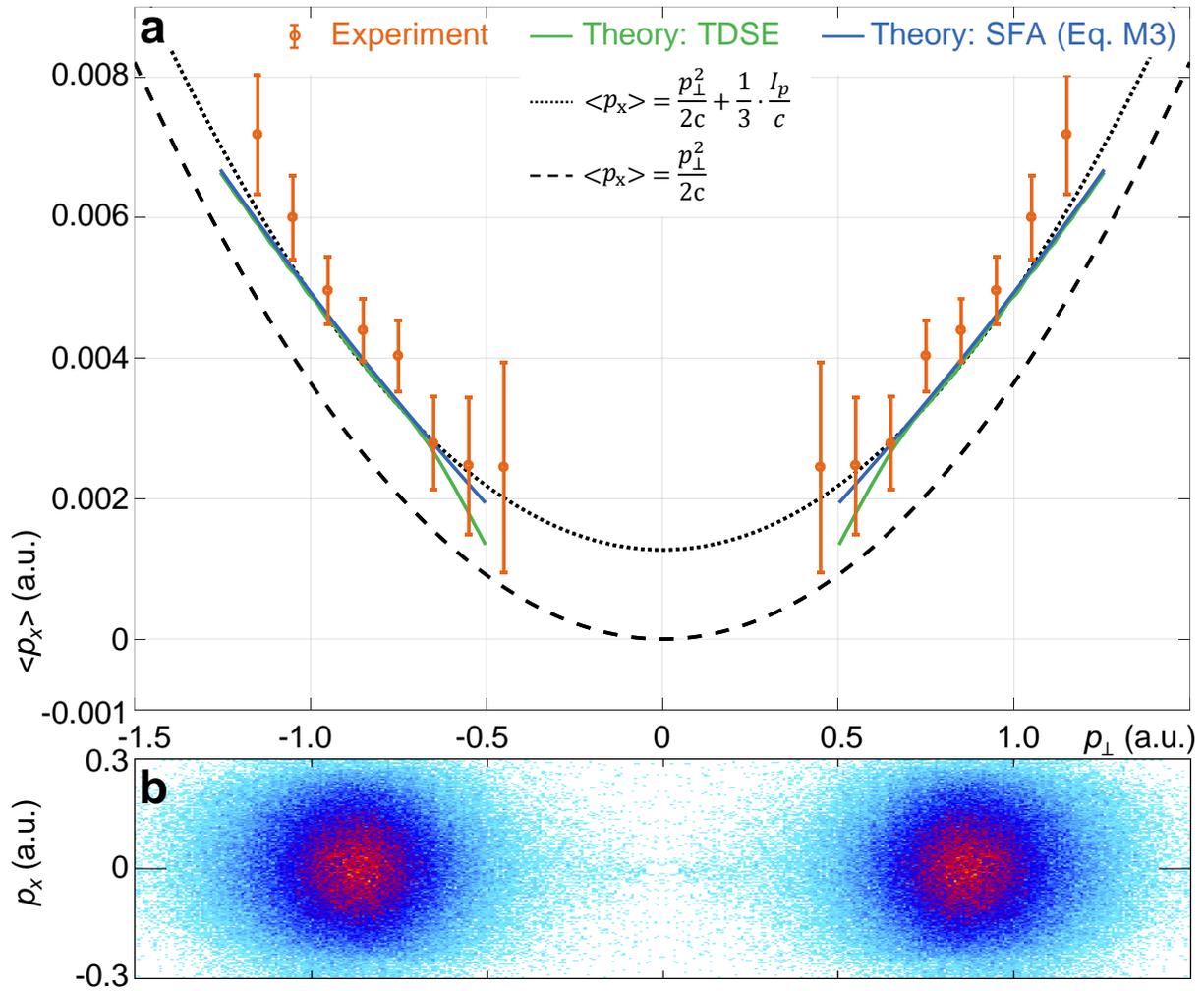

**Figure 2 | Results for circularly polarized light. a** The mean electron momentum in light propagation direction $\langle p_x \rangle$ is plotted against the radial momentum $|p_\perp| = \sqrt{p_y^2 + p_z^2}$ obtained by using circularly polarized light at $1.7 \times 10^{14}$ W/cm²; note the scale of $\sim 10^{-3}$ a.u. Error bars show statistical errors. **b** For visual orientation the experimental two-dimensional momentum distribution is displayed within the same coordinate frame as in **a**. There are $2.8 \times 10^5$ counts plotted with the same linear color scale, used in Fig.1.

The experimental results in Fig. 2b show the donut-shaped electron momentum distribution obtained by ionization with circularly polarized laser light. The light propagation is along the positive $p_x$ direction. Cylindrical coordinates are used, exploiting the rotational symmetry around $p_x$ by calculating $p_\perp = \sqrt{p_y^2 + p_z^2}$. On the scale of Fig. 2b, the non-dipole forward shift of the distribution is again not visible. The corresponding mean momentum <$p_x$> is shown in Fig. 2a on a magnified scale. The data are obtained by averaging over the two "single-way" measurements, see methods for details. This averaging eliminates potential instrumental asymmetries occurring in $p_x$. The data show a dependence of <$p_x$> on the radial momentum $p_\perp$ resulting from the radiation pressure on the electron. As this dependence is very closely described by equation (2) (dotted line in Fig. 2a), the data provides clear evidence for the forward shift of 1/3 $I_p$/c, verifying theoretical predictions[1,2,8]. This offset is, furthermore, a clear signature of the magnetic field effects occurring in the tunnel ionization step. It results from the action of the laser magnetic field onto the electron, while it is under the tunnel barrier. It is yet another counter-intuitive facet of the tunneling process, suggesting that during tunneling the electron is already partially decoupled from the nucleus and able to absorb 1/3 of the momentum of the photons needed to free the electron.

We support our experimental data by a numerical solution of the non-dipole three-dimensional time-dependent Schrödinger equation (TDSE). The so found dependence of the average momentum <$p_x$> on the radial momentum $p_\perp$ (green line in Fig. 2a) is in good agreement with the experimental data. A quantum-orbit model (blue line in Fig. 2a) derived from the strong-field approximation[3] (SFA) shows deviations from the simple parabola. These are caused by the initial velocity of the electron immediately after the tunneling process. However, for a vanishing initial velocity the expected shift of ($E_{e,kin}$+$I_p$/3)/c is reproduced. Details on the TDSE and SFA can be found in the methods.

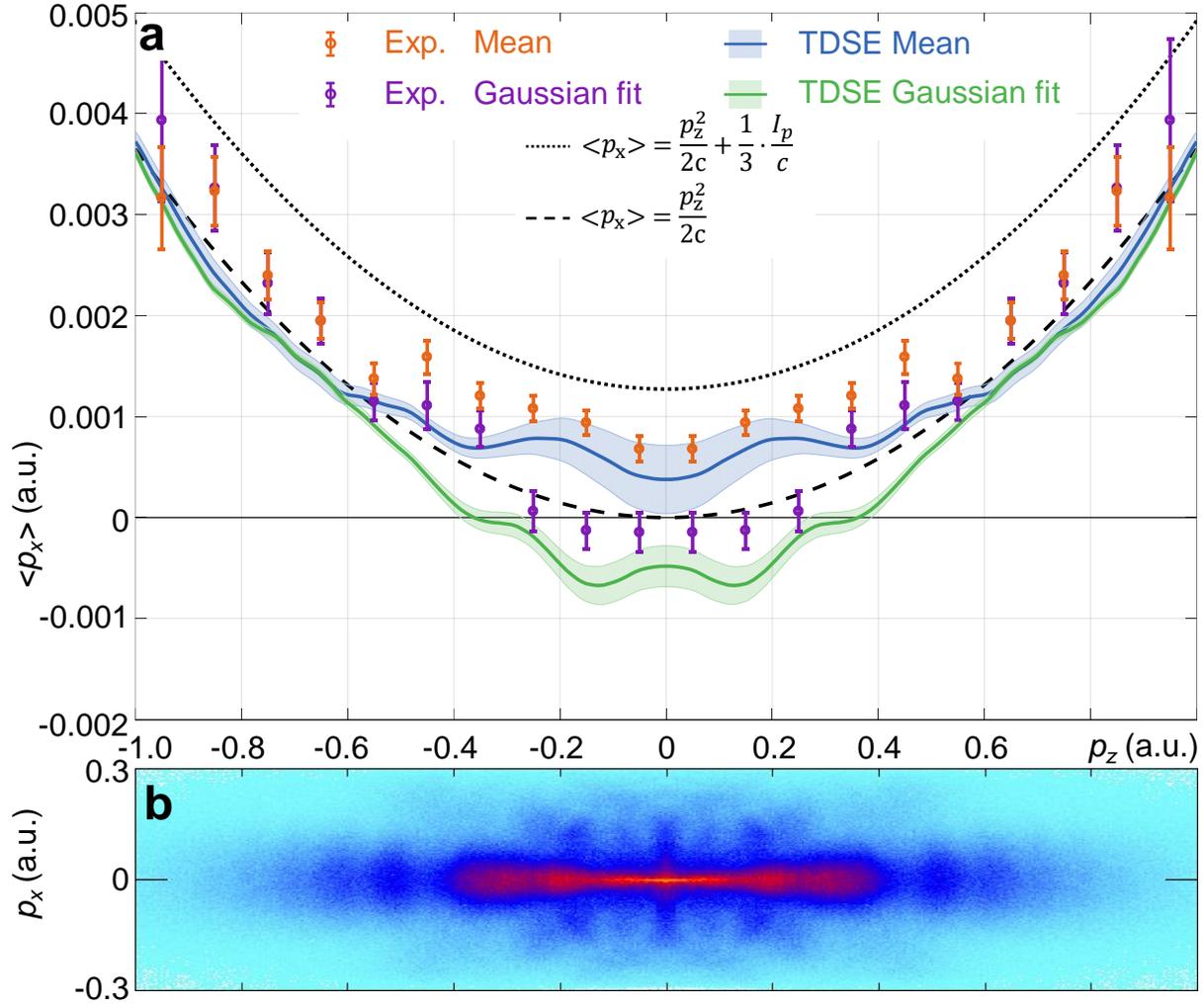

**Figure 3 | Results for linearly polarized light. a** The mean electron momentum in light propagation direction <$p_x$> is plotted against the momentum in direction of the light's polarization axis $p_z$ obtained by using linearly polarized light at $1.1 \times 10^{14}$ W/cm²; note the scale of ~ $10^{-3}$ a.u. Because of the symmetry of the experiment in light polarization axis z, the data was symmetrized with respect to $p_z = 0$ a.u. Experimental error bars show statistical errors. Theoretical error area is described in methods. **b** For visual orientation the experimental two-dimensional momentum distribution is displayed within the same coordinate frame as in **a**. There are $2.3 \times 10^6$ counts plotted with the same linear color scale, used in Fig.1.

Further aspects can be learned from the case of linear polarization where the freed electron can return to the parent ion after each half cycle of the laser pulse. This leads to strong post-tunneling Coulomb interaction and hence one expects deviations from equation (2). Fig. 3 shows experimental data ($<p_x>$, orange) along with the results obtained by solving the 3D TDSE (blue). The situation is more complex than for circularly polarized laser pulses as the mean $<p_x>$ does not coincide with the position of the peak of the momentum distribution, plotted as purple points (experiment) and green lines (TDSE) in Fig. 3a. To obtain the peak positions we performed Gaussian fits of the distributions (see methods for more details). The theoretical analysis shows that these positions are determined by rescattered electrons, which perform a "swing-by process" at the parent ion, as described in[6,7,25]. In the region of small momenta, the experimental ($|p_z| < 0.1$ a.u.) and more clearly the theoretical ($|p_z| < 0.36$ a.u.) Gaussian fit centers become negative, indicating a counter-intuitive "backward" shift of the peak of the momentum distribution against the laser propagation direction, as reported in previous experiments[6,7] and theory[4]. The electrons in the narrow central peak with nearly no kinetic energy are initially accelerated in the laser electric field with a slight forward push in laser direction by the magnetic field, but subsequently when returning to the parent ion scatter from it, resulting in an overall backward shift. At higher momenta ($|p_z| > 0.5$ a.u.), no significant $I_p$-dependent offset can be seen in our experiment and calculations for the case of linearly polarized light. In this region the position of the maximum is determined by the central fringe of the holographic pattern, that can be approximately described by glory rescattering[26]. Here, classical Coulomb focusing (as was recently described with the use of classical trajectory Monte Carlo simulations[25]) in combination with nearly constructive interference between different trajectories lead to the pronounced maximum. In contrast to the position of the maximum, the mean value $<p_x>$ is also influenced by "direct" (non-rescattered) electrons as well as the asymmetric emission strength in forward and backward direction. Taking these effects together, on average ($<p_x>$) the electron momentum behaves as one might expect from equation (1), i.e. showing an offset in light direction following radiation pressure.

Using two counter-propagating laser pulses, we present the first experimental results having sufficient accuracy to measure the influence of the laser magnetic field onto the electron in the classically forbidden tunnel barrier. The present study helps to fully understand strong-field ionization in a complete rigorous description including momentum conservation. For example we expect non-dipole modifications to the non-adiabatic momentum offsets[27]. The considered effects influence not only single electron emission to the continuum, but all strong field phenomena including non-sequential double ionization[28], high harmonic generation[14,15] and other ultrafast tools such as laser driven electron diffraction[11,12] and holography[13]. For future studies, we propose using two-color laser fields to significantly alter the magnetic field and thereby explore non-trivial photon momenta in the non-dipole regime of strong-field ionization. Besides modification of the light, modification of the target (i.e. usage of molecules) can illuminate the non-dipole response by the ionic core.

## Methods

### Experimental measurements

**Laser setup and gas target**

The laser pulses are generated by a *Coherent Legend Elite Duo* laser-system at a central wavelength of 800 nm. The repetition rate is 10 kHz and the duration of the laser pulses is 25 fs (full-width at half-maximum in intensity). The laser can be focused into the chamber from two opposite directions. To this end a dielectric beamsplitter, which is located 2.3 meters away from the entrance windows, splits the initial laser beam path and directs it into two pathways. Both pathways allow for the adjustment of intensity, polarization state and time delay of the laser pulses by neutral-density filters, $\lambda/2$ and $\lambda/4$ retardation plates and a delay stage, respectively. Both laser pulses are focused by identical lenses (f = 25 cm), which are placed outside of the vacuum chamber, into an argon gas target. The gas target is created by supersonic gas expansion from a small nozzle (opening hole diameter 30 μm) into vacuum and subsequently lead through a skimmer (diameter 210 μm). The transversal size of the gas jet can be adjusted by piezo-controlled collimators. In the measurements, the intersection of cropped gas jet and laser focus create an ionization volume of appr. 100×30×30 μm³, thereby reducing focal averaging. For the calibration of the laser intensity, the average drift momentum of the electrons ($p_{drift}$ = 0.87 a.u.), created by circularly polarized light, is used. For circularly polarized light, a laser pulse with a negative vector potential of 0.87 a.u. corresponds to an intensity of $1.7 \times 10^{14}$ W/cm² in the focus (Keldysh parameter $\gamma = 0.87$). The comparison of the laser powers used in the two polarization states yields an intensity of $1.1 \times 10^{14}$ W/cm² for linearly polarized light ($\gamma = 1.1$). We estimate an uncertainty of 10%.

**Particle detection & calibration**

Electrons and ions produced by strong-field ionization are guided towards position-sensitive detectors by a homogenous electric field (20 V/cm). For the electrons (ions) the length of the field region is 15 cm (58 cm), followed by a field-free region with a length of 30 cm (108 cm). The earth magnetic field is compensated by Helmholtz coils and the spectrometer is further shielded from remaining magnetic fields by a µ-metal shield. The signal of the electrons (ions) are amplified by a stack of 3 (2) multi-channel plates with a diameter of 80 mm (120 mm). Subsequently, for both types of particles a delay-line anode with 3 layers is used to measure the three-dimensional momentum. As is typical for a COLTRIMS reaction microscope[24] the momenta of electrons and ions are measured in coincidence. The momentum resolution is $\Delta p_x = \Delta p_y = 0.0011$ a.u., $\Delta p_z = 0.032$ a.u. for electrons (x points along the light-propagation direction, y points along the direction of the gas jet and z points along the time-of-flight direction). In previous experimental studies[5–7], the precise knowledge of the respective zero of the momentum distribution was obtained from electrons stemming from highly excited Rydberg states. These are created in the laser pulse by frustrated tunnel ionization[29,30] and are field ionized by the spectrometer field after the laser pulse has faded[31]. By assuming that the Rydberg-electrons experience no non-dipole effects through the excitation process and that a homogenous detection efficiency is given over the complete momentum distribution, the reported momentum offsets were determined. As described in the main text, in this study the zero-point for the electron momenta in x-direction is found by comparing the two "single-way" experiments. As an additional cross-check, the results obtained upon ionization by the standing wave are used, which should lie mirror-symmetrically around $p_x = 0$ a.u. The proportionality constant to obtain the correct final magnitude of electron momenta in x-direction and y-direction for the discussed small magnitudes is calibrated by analyzing a sharp momentum sphere ($|p| = 0.026$ a.u.) which is due to doubly excited auto-ionizing Rydberg-atoms in argon upon the irradiance with a strong linearly polarized light-field.[32] The proportionality constant to obtain the correct final magnitude of electron momenta in z-direction is calibrated with ATI-peaks from argon upon strong field ionization with linearly polarized light.

## Data analysis

In the case of circularly polarized light, the value of momentum in light propagation direction $p_x$ is plotted against $|p_\perp| = \sqrt{p_y^2 + p_z^2}$, see Fig. 2b. The three (see Fig. 1a-c) resulting two-dimensional distributions are sliced along $p_\perp$ with a binning of $p_\perp = 0.1$ a.u. For the resulting $p_x$-distributions the mean $<p_x>$ is calculated. The difference of the two mean values that stem from the two "single-way experiments" (see Fig. 1a+c) is divided by two. This procedure eliminates potential instrumental asymmetries. The result is shown in Fig. 2a. The error bars show the statistical error (68% confidence interval).

For linearly polarized light (polarization axis is aligned along the z-direction), an analogous analysis procedure is used. The 3D photoelectron momentum distributions (PMD) are projected on to the x-z plane in momentum space. These 2D distributions are sliced along $p_z$ with a binning of $p_z = 0.1$ a.u. For each emerging $p_x$-distribution the mean $p_x$-value is calculated and a Gaussian function is fitted to the central region of the distribution (range of region $|p_x| \leq 0.01$–$0.17$ a.u. dependent on the value of $p_z$). The mean values as well as the centers of the fitted Gaussian functions are plotted in Fig. 3, along with their respective statistical error (68% confidence interval).

## Numerical simulations

### Time-dependent Schrödinger equation (TDSE) simulations

The PMD are obtained by performing numerical simulations of the TDSE in the single-active electron approximation including leading-order non-dipole corrections. We follow the scheme presented in Ref.[21] such that the theory covers the dynamics within the electric quadrupole and magnetic dipole approximation. After applying a unitary transformation to the initial system in Coulomb gauge, we obtain a numerical solution of the TDSE, $i\partial_t \psi(\mathbf{r},t) = H\psi(\mathbf{r},t)$, with a transformed Hamiltonian

$$H = \frac{1}{2}\left(\mathbf{p} + \mathbf{A}(t) + \frac{e_z}{c}\left(\mathbf{p}\cdot\mathbf{A}(t) + \frac{1}{2}\mathbf{A}^2(t)\right)\right)^2 + V\left(\mathbf{r} - \frac{z}{c}\mathbf{A}(t)\right) \qquad (M1)$$

that is solved numerically using the split-operator method on a Cartesian grid with a time step of 0.025 a.u. While propagating until the final time, outgoing parts of the wave function are projected onto Volkov states and summed up coherently to obtain the momentum distribution[33]. The effective potential V for the argon atom is chosen as by Tong et al.[34], but with the singularity removed using a pseudopotential[35] for angular momentum l=1. The outermost subshell of the ground state consists of three degenerate p orbitals, $p_{+1}$, $p_{-1}$ and $p_0$, where the index indicates the magnetic quantum number $m_l$ of the orbital angular momentum in direction of light.

For circularly polarized light, we perform separately calculations for the $p_\pm$-states co- and counter-rotating with respect to the field as initial states. The used laser pulse has a carrier frequency $\omega = 0.0569$ a.u. and a sin²-envelope with full duration of 10 optical cycle. We perform calculations for 13 intensities ranging from $0.4 \times 10^{14}$ W/cm² to $1.6 \times 10^{14}$ W/cm² and average the results over the focal volume intensity distribution, assuming a Gaussian focus with a peak intensity of $I = 1.7 \times 10^{14}$ W/cm². The numerical grid size is 179 a.u. in all directions with a spacing of $\Delta x = 0.35$ a.u. The PMD is obtained with a resolution of $\Delta p_y = \Delta p_z = 0.0175$ a.u. and $\Delta p_x = 0.0088$ a.u. after propagating the wave function for one additional cycle after the end of the laser pulse. To calculate the average as a function of the in-plane momentum $p_\perp$ shown in Fig. 2 we follow the description from Ref.[21], but average over the ATI peaks as done in the experiment. Only minor changes on the level of $\Delta \langle p_x \rangle \approx 10^{-4}$ a.u. appear, when comparing the two p-states separately. Therefore, the incoherent sum of the momentum distributions is used to obtain Fig. 2.

For linearly polarized light, the $p_z$-state aligned along the polarization axis is irradiated with a 10-cycle sin²-envelope pulse of 0.0569 a.u. carrier frequency. The numerical grid size is 269 a.u. in all directions with a spacing of $\Delta x = 0.35$ a.u. The momentum distribution for a single peak intensity is obtained with a resolution of $\Delta p_y = \Delta p_z = 0.0116$ a.u. and $\Delta p_x = 0.0088$ a.u. after propagating the wave function for four additional cycles after the end of the pulse. We perform calculations for 15 intensities ranging from $0.4 \times 10^{14}$ W/cm² to $1.1 \times 10^{14}$ W/cm² and average the results over the focal volume intensity distribution, assuming a Gaussian focus with a peak intensity of $I = 1.1 \times 10^{14}$ W/cm². The PMD is also averaged over slices of $\Delta p_z = 0.1$ a.u. to reduce oscillations resulting from ATI rings. To determine the position of the lateral maximum we perform a Gaussian fit to the central region, as described in the experimental methods section, at each longitudinal momentum $p_z$. We find that there is only a minor difference in the position of the maximum obtained from projection onto the $p_x$-$p_z$-plane (as shown in Fig. 3) or the cut along this plane. The focal-averaged results show the same main features as the calculation for the highest intensity $1.1\times10^{14}$ W/cm², except that oscillations resulting from intra-cycle interferences are smoothed out and that the depth of the minima in the observables of Fig. 3 is slightly different. In addition to the 3D calculations in the main text, we check the convergence of our calculations and the stability of the extraction procedure by performing various calculations in 2D with different position and momentum grids. We estimate errors for both observables shown in Fig. 3 by using the maximal difference between the calculation with the highest resolution and calculations with the same grid parameters as in 3D but different extraction procedures. We find that for high momenta the results depicted in Fig. 3 are stable with respect to the discretization parameters. However, due to the rich low energetic structures and the appearance of Rydberg states, we find only qualitative accuracy at low energies.

**Quantum-orbit model based on strong-field approximation**

To deepen our understanding of the momentum transfer in recollision-free ionization with circularly polarized laser fields, we have developed a quantum-orbit model based on the strong-field approximation (SFA) as described in Ref.[3], but taking only first order corrections in $1/c$ into account. If we apply the saddle-point approximation to the SFA integral and neglect pre-exponential factors, we can solve the saddle-point equation for the complex-valued ionization time $t'_s$ exactly and write the photoelectron signal approximately as

$$w(\boldsymbol{p}) \approx e^{-2\mathrm{Im}S(\boldsymbol{p},t'_s)} \tag{M2}$$

with the imaginary part of the generalized action $\mathrm{Im}S(\boldsymbol{p}, t'_s)$. For fixed in-plane momentum $p_\perp = \sqrt{p_y^2 + p_z^2}$ the average $<p_x>(p_\perp)$ can be approximated by the shift of the maximum of the lateral distribution. In first order of $1/c$ we obtain for circular polarization that

$$p_x^{shift}(p_\perp) = \frac{1}{c}\left(\frac{p_\perp E_0}{\sqrt{2}\omega}\frac{\sqrt{\chi_\perp^2-1}}{\mathrm{acosh}(\chi_\perp)} - U_p\right) \tag{M3}$$

with 
$$\chi_\perp = \frac{\sqrt{2}\omega}{p_\perp E_0}\left(\frac{1}{2}p_\perp^2 + I_p + U_p\right) \tag{M4}$$

and the ponderomotive potential $U_p = E_0^2/(4\omega^2)$, where the expressions are written such that the actual electric field strength is $E_0/\sqrt{2}$. The analytical expression of equation (M3) is in quantitative agreement with the result calculated from the numerical solution of the TDSE for short-range potentials.

In the simplest possible picture, the photoelectrons can be described in a two-step model consisting of (i) laser-induced tunnel ionization and (ii) potential-free acceleration of the electron as a classical particle in the laser field. The acceleration maps an electron starting with a transverse initial velocity $\Delta v_0$ in the polarization plane to the final in-plane momentum component $p_\perp = \sqrt{2U_p} + \Delta v_0$. If we consider adiabatic tunneling, i.e. small Keldysh parameter $\gamma = \sqrt{2I_p}\,\omega/E_0$, and allow only for small initial velocities, the shift of equation (M3) can be approximated as

$$p_x^{shift}(p_\perp) \approx \frac{1}{c}\left(U_p + \sqrt{2U_p}\Delta v_0 + \frac{1}{3}\left(I_p + \frac{\Delta v_0^2}{2}\right)\right). \tag{M5}$$

The first two terms result from the potential-free acceleration and can be obtained by solving Newton's equation with an initial velocity. In contrast, the third term is attributed to the momentum transfer during the quantum mechanical under-the-barrier motion and hence cannot be interpreted classically. We want to emphasize that the initial velocity changes the momentum transfer mechanism compared to SFA for linear polarization such that the shift is not a simple quadratic function of the momentum $p_\perp$.

**Acknowledgements**

A.H., K.F. and K.H. acknowledge support by the German National Merit Foundation. We acknowledge support from Deutsche Forschungsgemeinschaft via Sonderforschungsbereich 1319 (ELCH) and by DFG Priority Programme "Quantum Dynamics in Tailored Intense Fields".

# References

1. Chelkowski, S., Bandrauk, A. D. & Corkum, P. B. Photon Momentum Sharing between an Electron and an Ion in Photoionization: From One-Photon (Photoelectric Effect) to Multiphoton Absorption. *Phys Rev Lett* **113**, 263005 (2014).

2. Klaiber, M., Yakaboylu, E., Bauke, H., Hatsagortsyan, K. Z. & Keitel, C. H. Under-the-Barrier Dynamics in Laser-Induced Relativistic Tunneling. *Phys Rev Lett* **110**, 153004 (2013).

3. He, P.-L., Lao, D. & He, F. Strong Field Theories beyond Dipole Approximations in Nonrelativistic Regimes. *Phys. Rev. Lett.* **118**, (2017).

4. Chelkowski, S., Bandrauk, A. D. & Corkum, P. B. Photon-momentum transfer in multiphoton ionization and in time-resolved holography with photoelectrons. *Phys Rev A* **92**, 051401 (2015).

5. Smeenk, C. T. L. *et al.* Partitioning of the Linear Photon Momentum in Multiphoton Ionization. *Phys Rev Lett* **106**, 193002 (2011).

6. Ludwig, A. *et al.* Breakdown of the Dipole Approximation in Strong-Field Ionization. *Phys Rev Lett* **113**, 243001 (2014).

7. Maurer, J. *et al.* Probing the ionization wave packet and recollision dynamics with an elliptically polarized strong laser field in the nondipole regime. *Phys Rev A* **97**, 013404 (2018).

8.  Yakaboylu, E., Klaiber, M., Bauke, H., Hatsagortsyan, K. & H. Keitel, C. Relativistic features and time delay of laser-induced tunnel-ionization. *Phys. Rev. A* **88**, 063421 (2013).

9.  Titi, A. S. & Drake, G. W. F. Quantum theory of longitudinal momentum transfer in above-threshold ionization. *Phys Rev A* **85**, 041404 (2012).

10. Reiss, H. R. Relativistic effects in nonrelativistic ionization. *Phys Rev A* **87**, 033421 (2013).

11. Meckel, M. *et al.* Laser-Induced Electron Tunneling and Diffraction. *Science* **320**, 1478–1482 (2008).

12. Blaga, C. I. *et al.* Imaging ultrafast molecular dynamics with laser-induced electron diffraction. *Nature* **483**, 194 (2012).

13. Huismans, Y. *et al.* Time-Resolved Holography with Photoelectrons. *Science* **331**, 61–64 (2011).

14. McPherson, A. *et al.* Studies of multiphoton production of vacuum-ultraviolet radiation in the rare gases. *J Opt Soc Am B* **4**, 595–601 (1987).

15. Ferray, M. *et al.* Multiple-harmonic conversion of 1064 nm radiation in rare gases. *J. Phys. B At. Mol. Opt. Phys.* **21**, L31 (1988).

16. Pisanty, E. *et al.* High harmonic interferometry of the Lorentz force in strong mid-infrared laser fields. *New J. Phys.* **20**, 053036 (2018).

17. Paul, P. M. *et al.* Observation of a Train of Attosecond Pulses from High Harmonic Generation. *Science* **292**, 1689–1692 (2001).

18. Baltuška, A. *et al.* Attosecond control of electronic processes by intense light fields. *Nature* **421**, 611 (2003).

19. Ivanov, I. A., Dubau, J. & Kim, K. T. Nondipole effects in strong-field ionization. *Phys Rev A* **94**, 033405 (2016).

20. Chelkowski, S., Bandrauk, A. D. & Corkum, P. B. Photon-momentum transfer in photoionization: From few photons to many. *Phys Rev A* **95**, 053402 (2017).

21. Brennecke, S. & Lein, M. High-order above-threshold ionization beyond the electric dipole approximation. *J. Phys. B At. Mol. Opt. Phys.* **51**, 094005 (2018).

22. Keil, T. & Bauer, D. Coulomb-corrected strong-field quantum trajectories beyond dipole approximation. *J. Phys. B At. Mol. Opt. Phys.* **50**, 194002 (2017).

23. Brennecke, S. & Lein, M. High-order above-threshold ionization beyond the electric dipole approximation: Dependence on the atomic and molecular structure. *Phys Rev A* **98**, 063414 (2018).

24. Dörner, R. *et al.* Cold Target Recoil Ion Momentum Spectroscopy: a 'momentum microscope' to view atomic collision dynamics. *Phys. Rep.* **330**, 95–192 (2000).

25. Daněk, J. *et al.* Interplay between Coulomb-focusing and non-dipole effects in strong-field ionization with elliptical polarization. *J. Phys. B At. Mol. Opt. Phys.* **51**, 114001 (2018).

26. Xia, Q. Z., Tao, J. F., Cai, J., Fu, L. B. & Liu, J. Quantum Interference of Glory Rescattering in Strong-Field Atomic Ionization. *Phys Rev Lett* **121**, 143201 (2018).

27. Eckart, S. *et al.* Direct Experimental Access to the Nonadiabatic Initial Momentum Offset upon Tunnel Ionization. *Phys Rev Lett* **121**, 163202 (2018).

28. Emmanouilidou, A., Meltzer, T. & Corkum, P. B. Non-dipole recollision-gated double ionization and observable effects. *J. Phys. B At. Mol. Opt. Phys.* **50**, 225602 (2017).

29. Nubbemeyer, T., Gorling, K., Saenz, A., Eichmann, U. & Sandner, W. Strong-Field Tunneling without Ionization. *Phys Rev Lett* **101**, 233001 (2008).

30. de Boer, M. P., Hoogenraad, J. H., Vrijen, R. B., Noordam, L. D. & Muller, H. G. Indications of high-intensity adiabatic stabilization in neon. *Phys Rev Lett* **71**, 3263–3266 (1993).

31. Diesen, E. *et al.* Dynamical Characteristics of Rydberg Electrons Released by a Weak Electric Field. *Phys Rev Lett* **116**, 143006 (2016).

32. Fechner, L. *et al.* Creation and survival of autoionizing states in strong laser fields. *Phys Rev A* **92**, 051403 (2015).

33. Lein, M., Gross, E. K. U. & Engel, V. Intense-Field Double Ionization of Helium: Identifying the Mechanism. *Phys Rev Lett* **85**, 4707–4710 (2000).

34. Tong, X. M. & Lin, C. D. Empirical formula for static field ionization rates of atoms and molecules by lasers in the barrier-suppression regime. *J. Phys. B At. Mol. Opt. Phys.* **38**, 2593–2600 (2005).

35. Troullier, N. & Martins, J. L. Efficient pseudopotentials for plane-wave calculations. *Phys Rev B* **43**, 1993–2006 (1991).